\pdfoutput=1 
\documentclass[aps,prd,twocolumn,superscriptaddress,tightelines,nofootinbib]{revtex4}
\usepackage{graphicx}
\usepackage{rotating}
\usepackage{epsfig}
\usepackage{bm}
\usepackage{latexsym,amssymb,amsmath,amsfonts,amssymb,txfonts,pxfonts,wasysym,float}
\usepackage{multirow}
\usepackage{color}
\newcommand{\beq}[1]{\begin{equation}\label{#1}}
\newcommand{\eeq}{\end{equation}}
\newcommand{\bea}[1]{\begin{eqnarray} \label{#1}}
\newcommand{\eea}{\end{eqnarray}}
\newcommand{\ba}{\begin{array}}
\newcommand{\ea}{\end{array}}

\def\be{\begin{equation}}
\def\ee{\end{equation}}
\def\gs{\mathrel{
   \rlap{\raise 0.511ex \hbox{$>$}}{\lower 0.511ex \hbox{$\sim$}}}}
\def\ls{\mathrel{
   \rlap{\raise 0.511ex \hbox{$<$}}{\lower 0.511ex \hbox{$\sim$}}}}

\newcommand{\postscript}[2]{\setlength{\epsfxsize}{#2\hsize}
   \centerline{\epsfbox{#1}}}
\newcommand{\e}{\varepsilon}

\newcommand{\comment}[1]{}

\usepackage[usenames,dvipsnames]{xcolor}
\definecolor{orange}{cmyk}{0,0.5,1,0}
\definecolor{rossoCP3}{cmyk}{0,.88,.77,.40}
\definecolor{graa}{rgb}{0.8,0.8,0.8}
\definecolor{blaa}{rgb}{0.2,0.2,0.6}

\begin{document}

\title{\color{rossoCP3}{The photo-disintegration of $^4$He on the
    cosmic microwave background \\ is less severe than earlier thought
}}

\author{Jorge F. Soriano}
\affiliation{Department of Physics \& Astronomy,  Lehman College, City University of
  New York, NY 10468, USA
}
\affiliation{Department of Physics,
 Graduate Center, City University
  of New York,  NY 10016, USA
}

\author{Luis A. Anchordoqui}
\affiliation{Department of Physics \& Astronomy,  Lehman College, City University of
  New York, NY 10468, USA
}
\affiliation{Department of Physics,
 Graduate Center, City University
  of New York,  NY 10016, USA
}

\affiliation{Department of Astrophysics,
 American Museum of Natural History, NY
 10024, USA
}

\author{Diego F. Torres}
\affiliation{Institute of Space Sciences (IEEC-CSIC),  Campus UAB,
  Carrer de Magrans s/n, 08193 Barcelona, Spain 
}
\affiliation{Instituci\'o Catalana de Recerca i Estudis Avan\c{c}ats
  (ICREA),  E-08010 Barcelona, Spain
}

\affiliation{Institut d'Estudis Espacials de Catalunya (IEEC),
08034 Barcelona, Spain
}

\begin{abstract}
  \noindent We thoroughly study the photo-disintegration of {$^4$He}
  on the cosmic microwave background using the most recent
  cross-section data both from the inclusive measurement observing the
  analog of the giant dipole resonance in $^4$He through the
  charge-exchange spin-flip {$^4$He} ($^7$Li,$^7$Be) reaction and from
  measurements of exclusive two-body and three-body processes:
  ${^4{\rm He}} \, (\gamma,p)\, {^3{\rm H}}$, ${^4{\rm He}} \,
  (\gamma,n) \, {^3{\rm He}}$, and {$^4$He} $(\gamma, pn)$ {$^2$H}. We
  show that the present-day (redshift $z=0$) mean free path of
  ultra-relativistic (Lorentz factor $\sim 10^{10}$) helium nuclei
  increases by more that 15\% with respect to previous estimates
  adopted as benchmarks for Monte Carlo simulation codes of
  ultrahigh-energy cosmic ray propagation.  This implies that the {\it physical}
  survival probability of $^4$He nuclei would be larger than predicted
  by existing event generators.  For example, for $E \sim
  10^{10.8}~{\rm GeV}$ and a propagation distance of 3.5~Mpc, the
   $^4$He intensity would be $35\%$ larger than the output of
  CRPropa 3 program and $42\%$ larger than the output of SimProp v2r4
  program. We provide new parametrizations for the two-body and
  three-body photo-disintegration cross-sections of $^4$He, $^3$He,
  tritium, and deuterium.
\end{abstract}

\maketitle

\section{Introduction}

The Greisen-Zatsepin-Kuzmin (GZK) horizon of
helium~\cite{Greisen:1966jv,Zatsepin:1966jv} is a key parameter in
ascertaining the contribution of ultrahigh-energy ($E \agt
10^{10}~{\rm GeV}$) cosmic rays (UHECRs) with directional pointing to
nearby sources. Numerical~\cite{Allard:2008gj,Allard:2011aa} and
analytical~\cite{Anchordoqui:2017abg} estimates of this parameter, as
well as Monte Carlo simulation codes of UHECR
propagation~\cite{Aloisio:2012wj,Kampert:2012fi,Batista:2015mea,Batista:2016yrx,Boncioli:2016lkt,Aloisio:2017iyh}
are customarily based on
fits~\cite{Puget:1976nz,Karakula:1993he,Rachen,Stecker:1998ib} to
cross-section measurements from the
sixties~\cite{Gorbunov:1,Gorbunov:2,Gorbunov:3,Gorbunov:4,Gorbunov:1968quk,Fuller},
which do not allow a precise description of the giant dipole resonance
(GDR) near threshold.

The first simultaneous measurement of the two-body and three-body
photo-disintegration cross-sections of $^4$He in the GDR region was
carried out in 2005 at the National Institute of Advanced Industrial
Science and Technology (AIST)~\cite{Shima:2005ix}. Data from the
three-body process yield a {$^4$He} ($\gamma,pn$) {$^2$H} cross
section of $0.04\pm 0.01$~mb at 29.8~MeV, in agreement with previous
measurements~\cite{Gorbunov:1,Gorbunov:2,Gorbunov:3,Arkatov:1970yd,Balestra:1979eq}. However, the
dominant $^4{\rm He} \, (\gamma,p)\, {^3{\rm H}}$ and ${^4{\rm He}} \,
(\gamma,n) \, {^3{\rm He}}$ cross sections are found to increase
monotonically with energy up to 29.8~MeV, in strong disagreement with
previous
observations~\cite{Berman:1980zz,Calarco:1983zz,Bernabei:1988rq,Nilsson:2005ah}. Subsequently,
a detailed studied of the GDR in $^4$He was carried out at the
Research Center for Nuclear Physics (RCNP), using a 455~MeV
$^7$Li$^{3+}$ beam bombarding a $^4$He gas target cooled to about
10~K~\cite{Nakayama:2007zza,Nakayama:2008zza}.  An indirect
measurement of the GDR in {$^4$He} was obtained by observing its
analog via the $^4$He ($^7$Li, $^7$Be) reaction at forward scattering
angles. The inclusive cross-section measurement from the $^4$He
($^7$Li,$^7$Be) reaction also shows a radical departure from the
results of the AIST group. Deepening the mystery, the $^4$He
photo-disintegration cross section was measured again by the same group at
AIST, confirming their earlier findings~\cite{Shima:2010hno}. To clarify the
situation, the total (i.e. angle-integrated) cross-section of the
exclusive two-body channels was measured at the High Intensity
Gamma-ray Source (HI$\gamma$S)~\cite{Raut:2012zz,Tornow:2012zz}.  The
HI$\gamma$S experiment confirmed that the peak of the GDR is near 
27~MeV and emphasized the differences with the AIST 
measurements. If we would assume that a systematic effect affected the AIST measurement 
of $^4{\rm He} \,
(\gamma,p)\, {^3{\rm H}}$ and ${^4{\rm He}} \, (\gamma,n) \, {^3{\rm
    He}}$ and leave these aside, we may conclude that there is now a
good agreement in the experimental front (see the data plotted in Fig. 1, top panel).

In this article we provide a
new parametrization of the photo-disintegration cross-section of
helium through a fit to the most recent data from the RCNP and
HI$\gamma$S experiments. Armed with this parametrization we re-examine
the opacity of the cosmic microwave background (CMB) to
ultra-relativistic (Lorentz factor $\sim 10^{10}$) helium nuclei.

\section{New parametrization of the GDR in $\bm{^4}$H\lowercase{e}}

The photo-absorption cross-section of a nucleus of charge $Ze$ and
baryon number $A$ roughly obeys the Thomas-Reiche-Kuhn (TRK) dipole sum
rule~\cite{Thomas,Reiche,Kuhn}
\begin{equation}
\Sigma = \int_0^\infty \sigma_A(\varepsilon) \ d \varepsilon = 59.8 \
\frac{Z \ (A-Z)}{A}~{\rm MeV \, mb}\, .
\end{equation}
Symmetric resonant cross-sections are commonly fitted by the normal
distribution, with probability density function given by
\begin{equation}
f_\mathcal N(\mu,\Gamma;\varepsilon) \equiv  
\frac{1}{\sqrt{2\pi\Gamma^2}}\exp\left[{-\frac{(\varepsilon-\mu)^2}{2\Gamma^2}}\right],
\label{eq:norm}\end{equation}
where $\mu$ is the mean and $\Gamma$
measures the dispersion around the mean.

The features of the cross-section data of the nuclei analyzed herein make evident
that the GDR does not follow a symmetric curve around its central value.
A simple way to account for the antisymmetry when the fall on the
right side of the central value is much slower than the rise on the
left side is to consider logarithmic distributions. These can be
obtained as $g(x)\ dx=f(\ln x) \ d\ln x$, if $f$ is a symmetric distribution, which gives $g(x)=f(\ln
x)/x$ for $x>0$. To accommodate threshold effects we can simply shift the
independent variable so that the threshold is at some value $x_{\rm th}$
rather than $0$.

To model the shape of the  photo-disintegration cross-section in the
energy range of the GDR  we adopt
the shifted log-normal distribution. Substituting 
$\e$ for $\ln(\e-\e_{\rm th})$ in (\ref{eq:norm})
and introducing the $1/(\e-\e_{\rm th})$ factor, we arrive
at the cross-section density function
\begin{equation}
 \sigma_A (\sigma_0,\e_0,\e_{\rm th},\Gamma;\e) = \sigma_0
\exp\left[-\frac{\ln^2\left(\frac{\e-\e_{\rm th}}{\e_0-\e_{\rm th}}\right)}{2\Gamma^2}\right]
\,,
\label{eq:shiftlognorm}
\end{equation}
where $\e_0$ is the central value of the GDR energy band (with
threshold $\e_{\rm th}$), $\sigma_0$
is the cross section at $\varepsilon=\varepsilon_0$, and $\Gamma$
is a measurement of the dispersion around $\e_0$.

For analytical order of magnitude estimates, it is convenient to
obtain a form of the cross-section in the single pole of the
narrow-width approximation (NWA). Introducing the change of variables
\begin{equation}
z(\e)\equiv\ln \left(\frac{\e-\e_{\rm th}}{\e_0-\e_{\rm th}} \right)
\end{equation}
we have
\begin{equation}
 \sigma_A (\sigma_0,\e_0,\e_{\rm th},\Gamma;\e)\propto f_\mathcal N
 (0,\Gamma;z(\e)) \, .
\end{equation} 
For the normal distribution, 
\begin{eqnarray}
\lim_{\Gamma\to0}f_\mathcal
  N(0,\Gamma;z(\e))  &  = & 
  \delta(z(\e))=\frac{\delta(\e-\e_0)}{|z'(\e_0)|}  \nonumber \\
& = &  (\e_0-\e_{\rm th}) \ \ \delta(\e-\e_0) \, ,
\end{eqnarray} 
and so we can approximate (\ref{eq:shiftlognorm}) as 
\begin{equation}
  \sigma_A (\sigma_0,\e_0,\e_{\rm th},\Gamma;\e)\approx {\cal A} \
  \delta(\e-\e_0),
\label{sigma}
\end{equation} 
where ${\cal A}$ is the normalization constant satisfying
\begin{equation}\int_{\e_{\rm th}}^\infty {\cal A} \ \delta(\e-\e_0) \
  d\e=\int_{\e_{\rm th}}^\infty \sigma_A(\sigma_0,\e_0,\e_{\rm th},\Gamma;\e) \
  d\e,\end{equation} 
and therefore
 \begin{equation}
{\cal A} =  \sqrt{2\pi} \ \sigma_0 \
\Gamma \ (\e_0-\e_{\rm th}) \ e^{\Gamma^2/2} \, .
\end{equation}

\begin{figure}[tbp] 
    \postscript{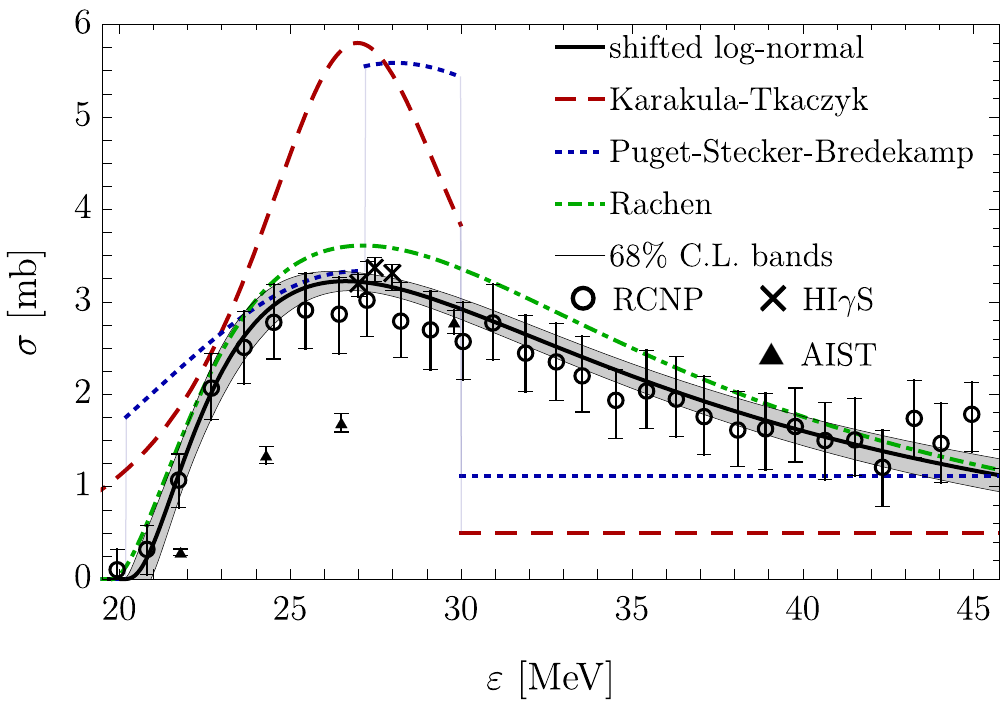}{0.9} 
    \postscript{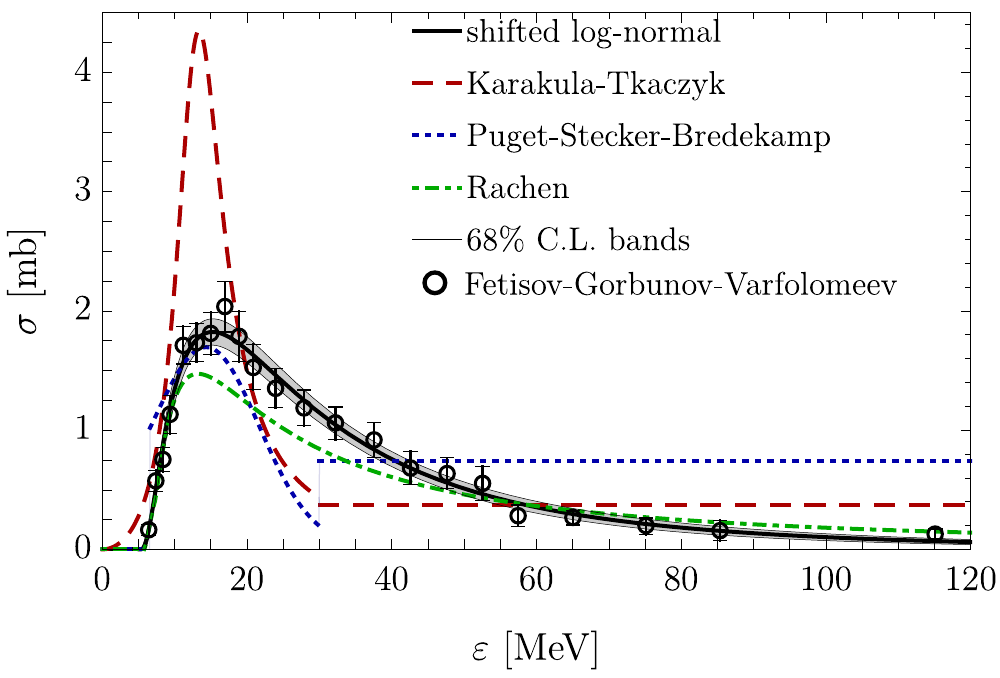}{0.9} 
  \postscript{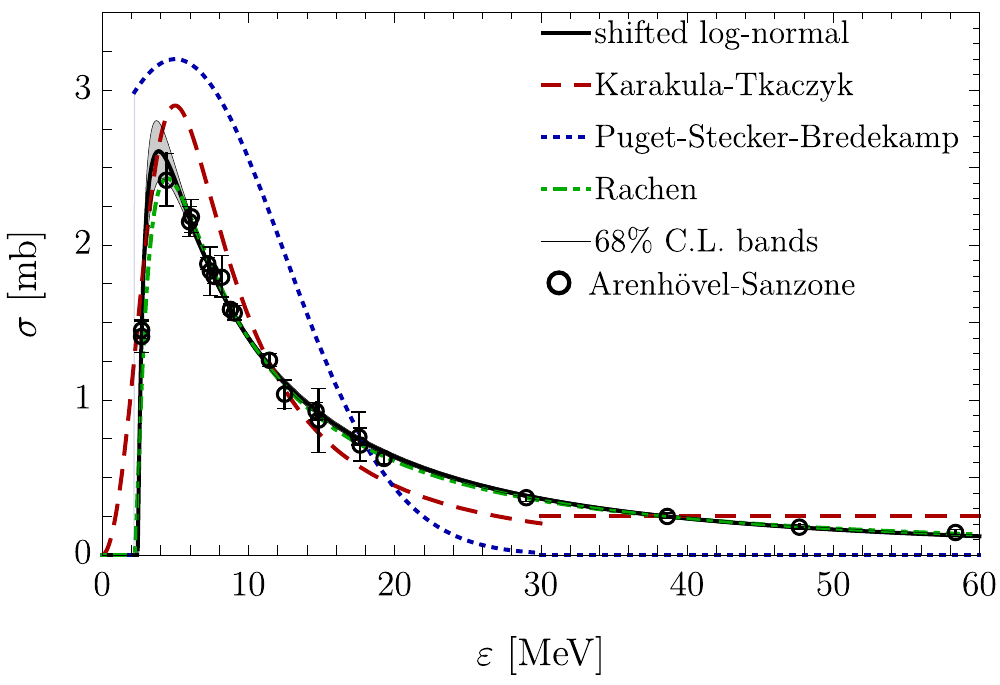}{0.9} 
\caption{Best fit and 68\% CL bands of the  $^4$He (top), $^3$He
  (middle), and $^{2}$H (bottom) 
  photo-disintegration cross section. Previous parametrizations of the
  cross section are also shown for visual comparison; for details one
  can refer to Appendix~\ref{app:1}. The experimental data have been
  taken from~\cite{Shima:2005ix,Nakayama:2007zza,Raut:2012zz,Tornow:2012zz} (top),
  \cite{Fetisov} (middle), and~\cite{Arenhovel:1990yg,Bacca:2014tla}
  (bottom).}  \label{fig:1} 
\end{figure}

Fitting (\ref{eq:shiftlognorm}) to the $^4$He data we find the four
parameters and corresponding 68\% C.L. band. The cross section
parameters are given in Table~\ref{tabla1} and shown in
Fig.~\ref{fig:1}.  For completeness, we also studied the
photo-disintegration of secondary $^3$He and $^2$H. The cross section
parameters are also given in Table~\ref{tabla1} and shown in
Fig.~\ref{fig:1}.  A comparison of our results with previous
approximations (which are briefly summarized in Appendix~\ref{app:1}) is also exhibited in Fig.~\ref{fig:1}. The
$^3$He and $^3$H (tritium) have similar photo-disintegration
properties. Any possible distinction because of the differences in
binding energy due to the Coulomb field disparity would fall within
theoretical and experimental uncertainties~\cite{Bacca:2014tla}.

\begin{table*}
\caption{Parameters of the photo-disintegration cross-section. \label{tabla1}}
\begin{tabular}{cccccc}
\hline
\hline
~~~~~~~~~$A$~~~~~~~~~&~~~~~~~~~$\sigma_0$ (mb)~~~~~~~~~&~~~~~~~~~$\e_0$ (MeV)~~~~~~~~~&~~~~~~~~~$\e_{\rm th}$
(MeV)~~~~~~~~~&~~~~~~~~~$\Gamma$~~~~~~~~~ &~~~~~~~~~${\cal A}$ (mb MeV)~~~~~~~~~ \\
\hline
4 & $3.22\pm 0.05$ & $26.6 \pm 0.4$ & $20.1 \pm 0.4$ & $0.94 \pm 0.08$
& $77\pm3 \phantom{0~}$
\\
3 & $1.82 \pm 0.05$ & $15.3 \pm 0.4$ & $5.1 \pm 0.2$ & $0.93 \pm 0.04$
& $67\pm 2\phantom{0~}$ \\
2 & $2.60 \pm 0.09$ & $3.87 \pm 0.09$ & $2.42 \pm 0.05$ & $1.48 \pm
0.04$ & $42.2 \pm 0.4$ \\
\hline
\hline
\end{tabular}
\end{table*}

\begin{figure*}[tbp] 
\begin{minipage}[t]{0.49\textwidth}
    \postscript{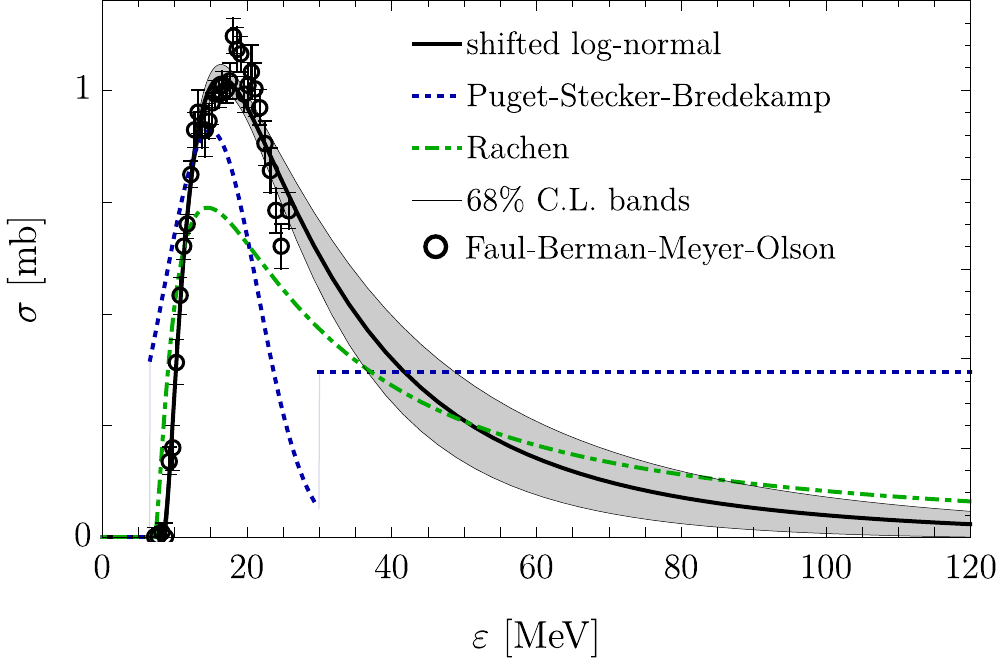}{0.99} 
\end{minipage} 
\begin{minipage}[t]{0.49\textwidth}
    \postscript{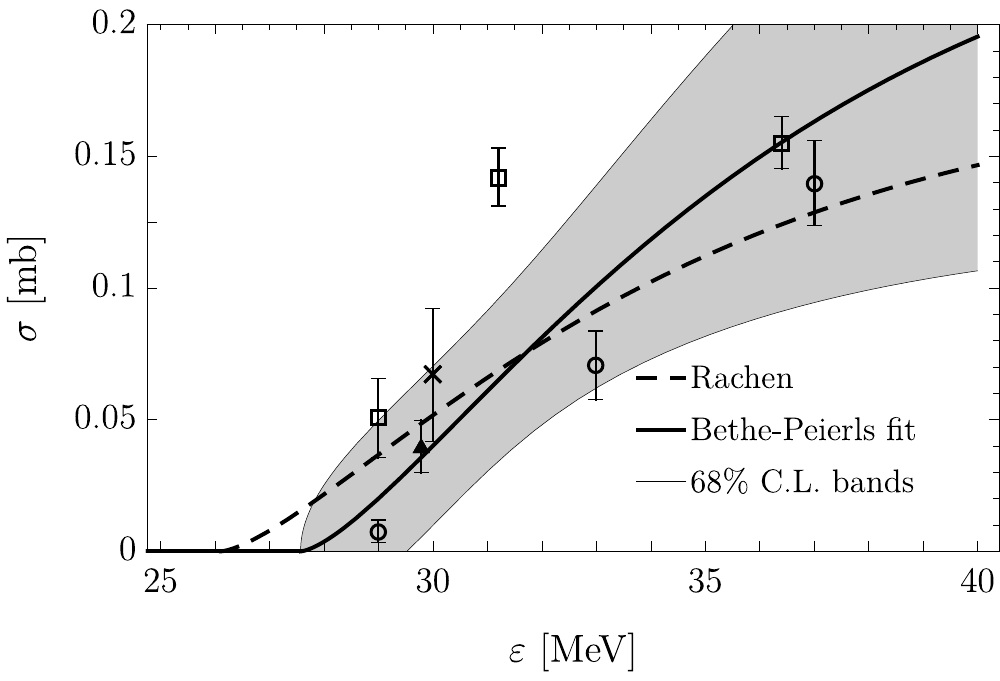}{0.99} 
\end{minipage} 
\caption{Best fit and 68\% CL bands of the exclusive  {$^3$He}
  $(\gamma, pn)$ {H} (left) and  {$^4$He}
    $(\gamma, pn)$ {$^2$H} (right)
  photo-disintegration cross section. Previous parametrizations of the
  cross section are also shown for visual comparison. The experimental data have been
  taken from~\cite{Faul:1981zz,Skibinski:2003de} (left) and $\blacktriangle$~\cite{Shima:2005ix},
    $\circ$~\cite{Arkatov:1970yd},
    $\times$~\cite{Gorbunov:1,Gorbunov:2,Gorbunov:3}, and
    $\Box$~\cite{Balestra:1979eq} (right).}  
\label{fig:2}
\end{figure*}

To complete our analysis of the photo-disintegration of light nuclei,
we provide the relevant branching ratios via fits to the cross-sections for the
exclusive three-body processes {$^3$He} $(\gamma, pn)$ {H} and
{$^4$He} $(\gamma, pn)$ {$^2$H}. The former can be modeled with a
shifted log-normal distribution, the best fit parameters are:
$\varepsilon_0 = (16.5 \pm 0.2)~{\rm MeV}$, $\Gamma = 0.97 \pm 0.07$,
$\varepsilon_{th} = (8.1 \pm 0.3)~{\rm MeV}$, and $\sigma_0= (1.03 \pm
0.01)~{\rm mb}$. The latter is best represented by a Bethe-Peierls
(BP) form~\cite{Bethe}, 
\begin{equation}
\sigma (\beta, B; \e) = \beta \times \sigma_{\rm{BP}}(B;\varepsilon)=
\beta \times \frac{\sigma_{{\rm T}p}}{\alpha_{\rm
      EM}}\frac{m_p
    c^2}{B}\frac{(x-1)^{3/2}}{x^3} \,,\end{equation}
with best fit parameters $\beta = 2.1 \pm 0.5$ and $B = 27.6 \pm 0.7~{\rm
  MeV}$. 
Here, $x = \e/B$, $\alpha_{\rm EM}$ is the fine structure constant, and $\sigma_{{\rm T}p}$ the Thomson cross section for the proton \begin{equation}\sigma_{{\rm T}p}=\frac{8\pi}{3}\left(\frac{\alpha_{\rm EM}\hbar c}{m_p c^2}\right)^2.\end{equation}
In Fig.~\ref{fig:2}  we show a comparison
of the best fit and 68\% C.L. bands for the cross sections of
three-body processes and previous estimates. To a good approximation, the ratio of the
photo-proton $^4$He ($\gamma,p$) $^3$H to the photo-neutron
{$^4$He} ($\gamma,n$) {$^3$He} cross
sections can be set equal to
one~\cite{Bernabei:1988rq,Raut:2012zz,Tornow:2012zz}.

In an aside, it is interesting to note that the Rachen's parameterization is the closest from all other earlier descriptions  
to the new experimental data. 
For instance, Rachen's description (developed $\sim 20$ years ago) would produce a higher cross section for  $^4$He, what at the end is critical for producing the 
effect on enlarging the propagation distance that we uncover below, but the general shape is quite 
acceptable. Something similar happens for   $^3$He and   $^2$H.

\begin{figure*}[tbp] 
\begin{minipage}[t]{0.49\textwidth}
    \postscript{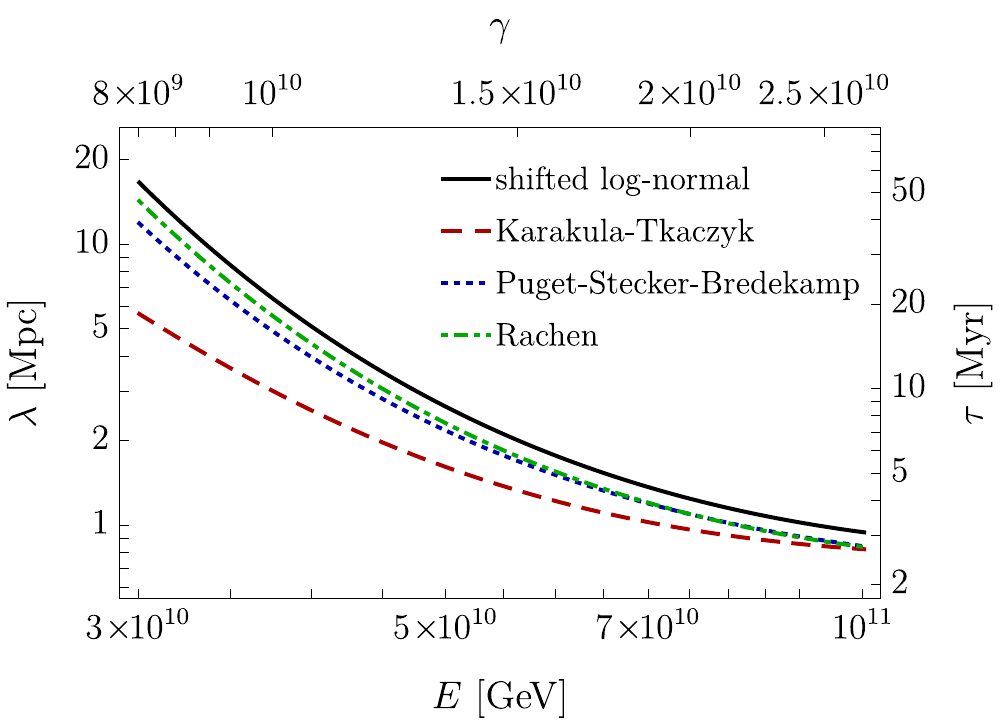}{0.99} 
\end{minipage} 
\begin{minipage}[t]{0.49\textwidth}
    \postscript{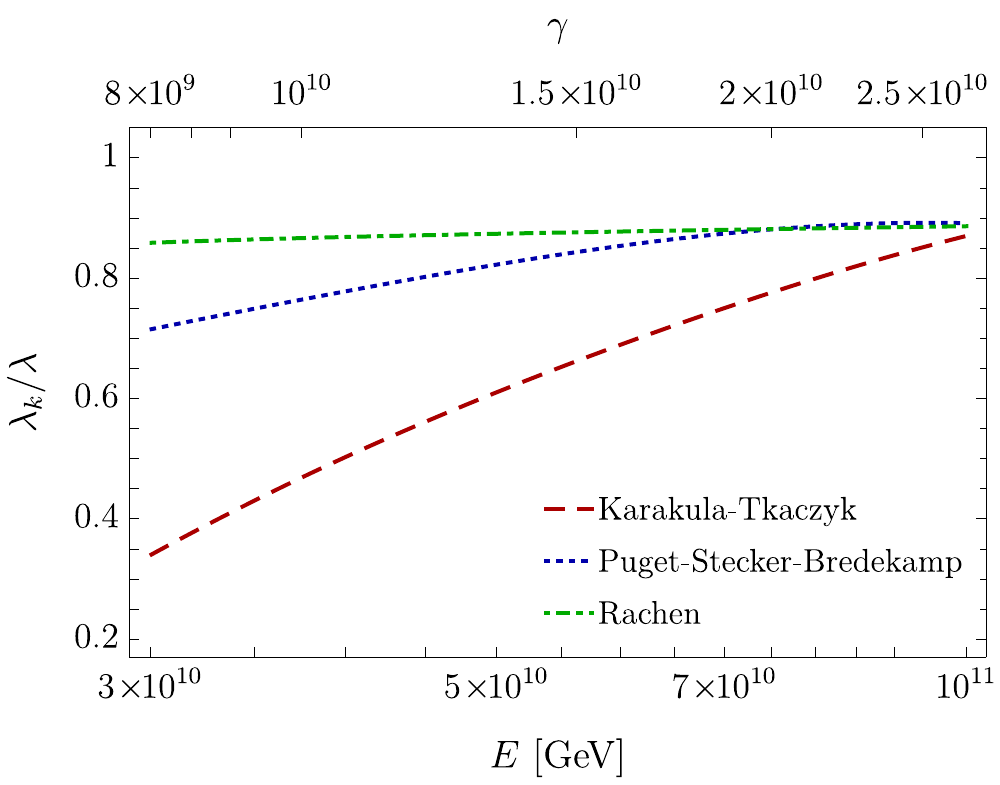}{0.9} 
\end{minipage} 
\caption{{\it Left:} Comparison of the various estimates of the mean free
  path of UHECR $^4$He nuclei propagating through the CMB at
  $z=0$ (left), and $\lambda_k(\gamma)/\lambda (\gamma)$, for $k\in\{{\rm
  KT, PSB, R}\}$ (right).} 
\label{fig:3}
\end{figure*}

\begin{figure}
  \postscript{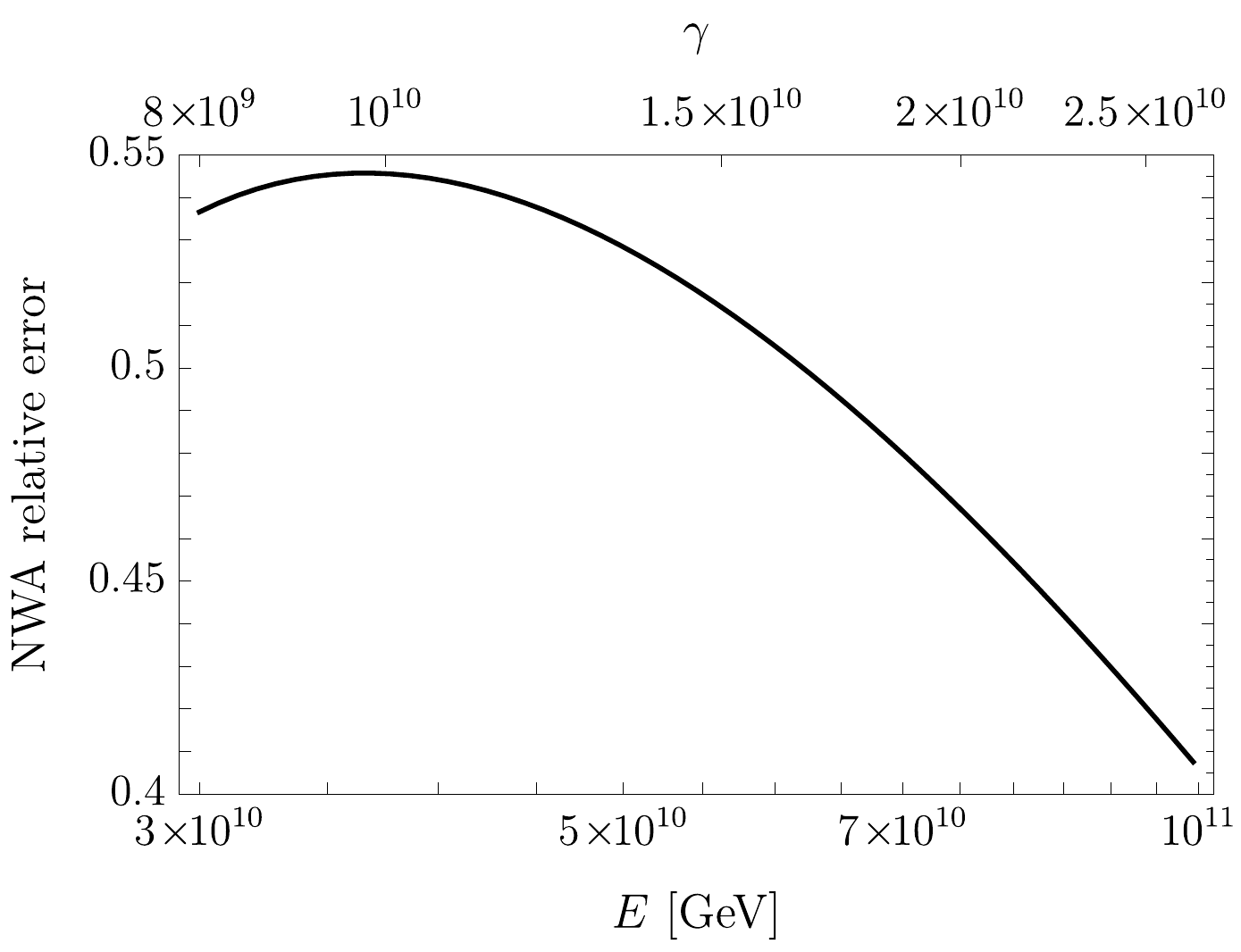}{0.9}  
\caption{Relative error $(\lambda-\lambda_{\rm NWA})/\lambda$ of the NWA.}  
\label{fig:4}
\end{figure}

\begin{figure}[tbp] 
    \postscript{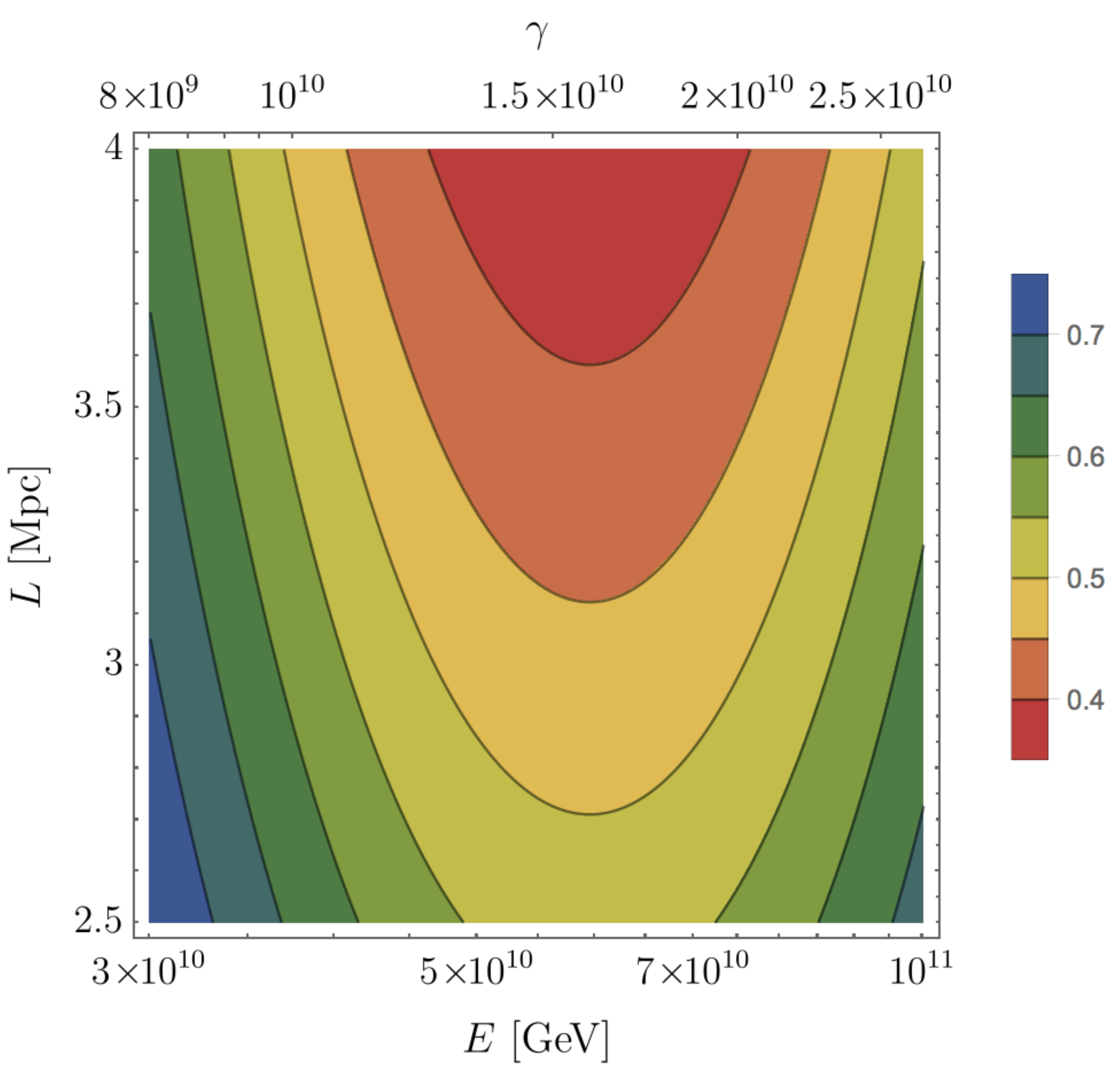}{0.99} 
    \postscript{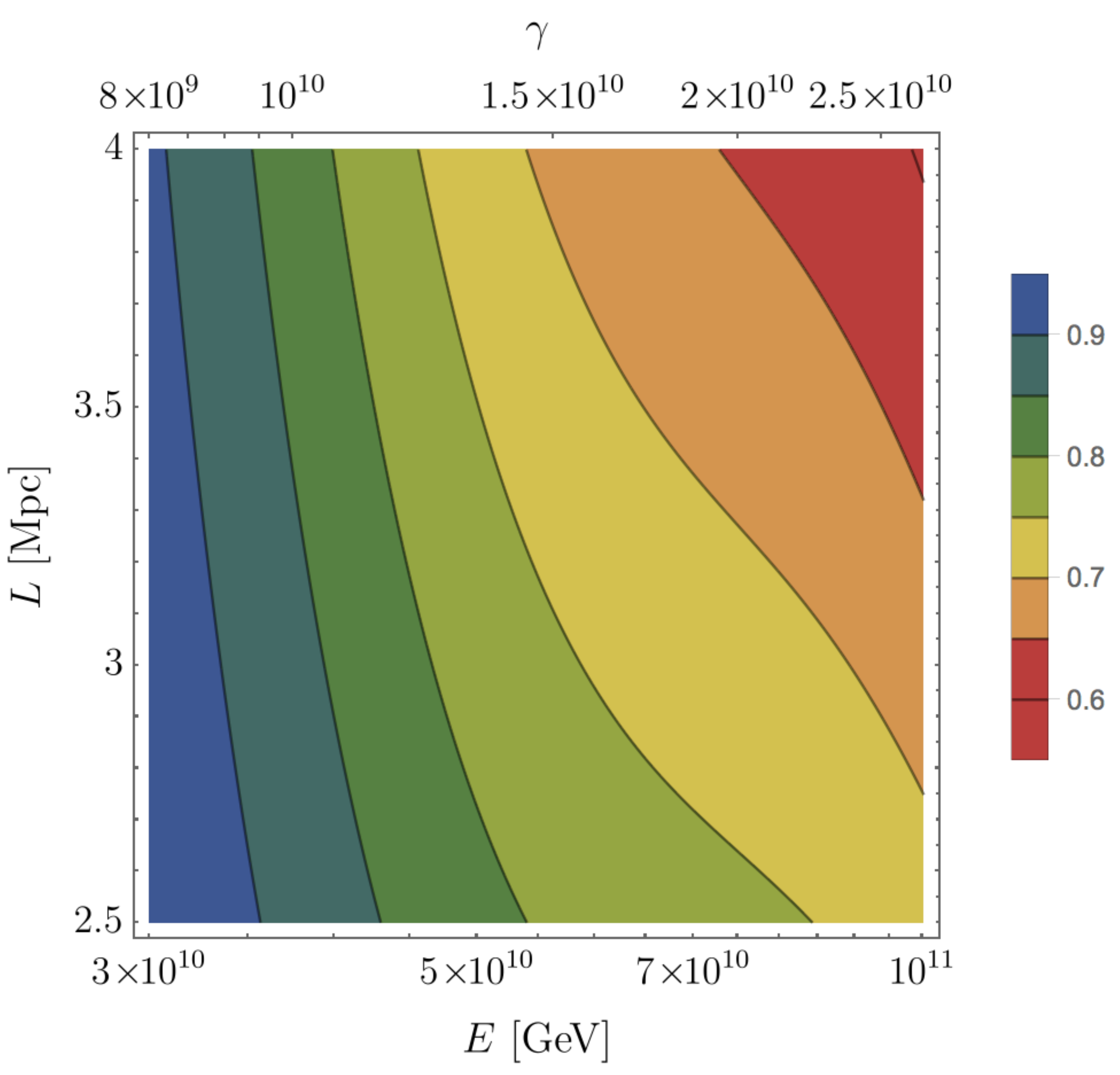}{0.99} 
  \postscript{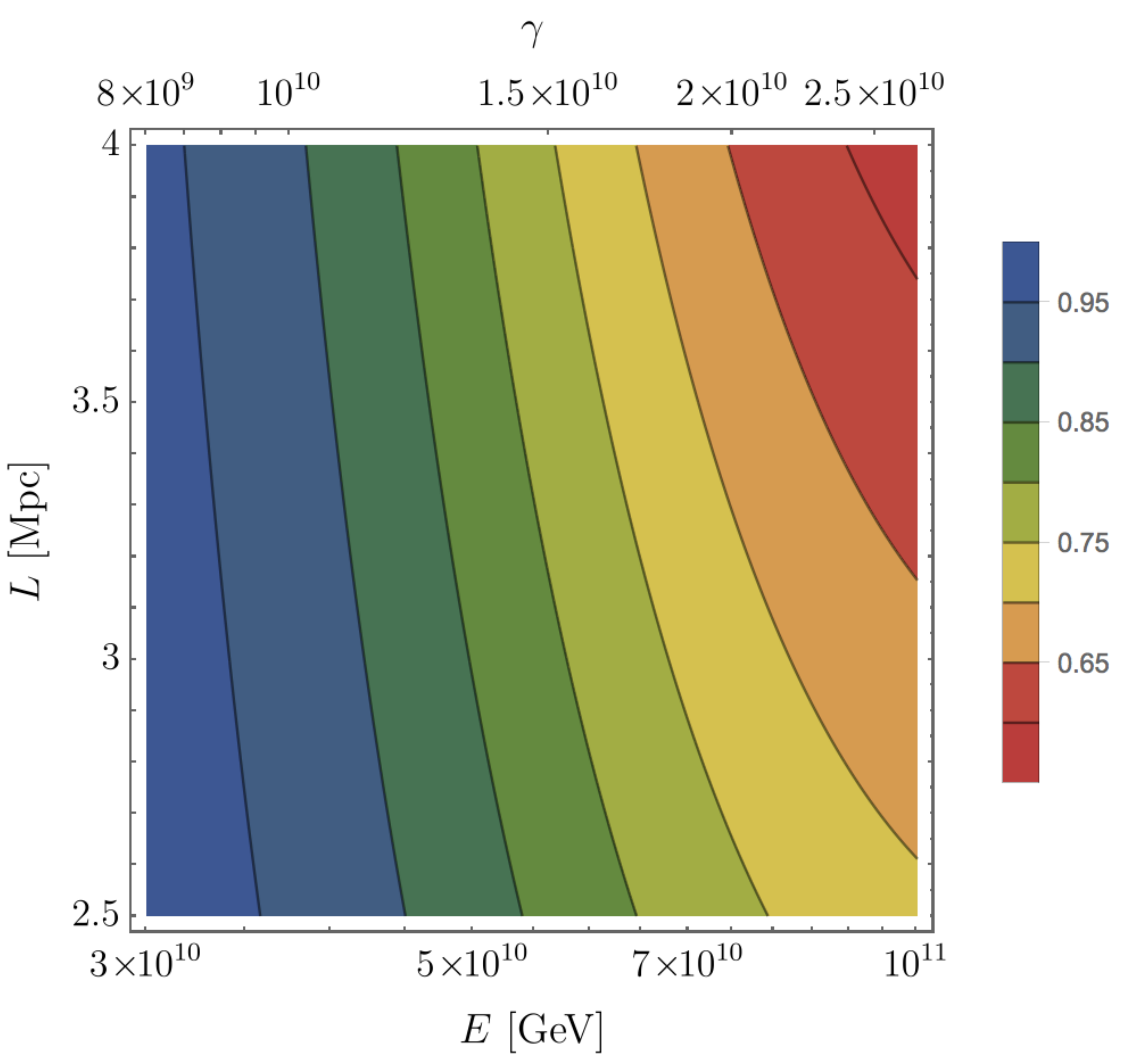}{0.99} 
\caption{Relative transmittance for $k = {\rm KT}$ (top), $k= {\rm
    PSB}$ medium, and $k = {\rm R}$ (bottom).}  
\label{fig:5} 
\end{figure}

\section{Photo-disintegration of $\bm{^4}$H\lowercase{e} on the CMB}

 We now turn to estimate the GZK energy loss of
ultra-relativistic $^4$He nuclei scattering off the CMB.  The relevant
mechanisms for the GZK energy loss of UHECR $^4$He nuclei are: {\it
  (i)}~$e^+ e^-$ pair production in the field of the nucleus, {\it
  (ii)}~photo-disintegration, and {\it (iii)}~photo-pion
production. In the nucleus rest-frame, pair production has a threshold
at $\sim 1~{\rm MeV}$. The inelasticity of pair production is very low
($\sim m_e/m_p$, for protons), so that the characteristic time-scale
of energy loss for this process at energies $E \agt 10^{10}~{\rm GeV}$
is $E/(dE/dt) \approx 10^{\rm 9.7}~{\rm
  yr}$~\cite{Aharonian:1994nn}. For a nucleus, the
energy loss rate is $Z^2/A$ times higher than for a proton of the same
Lorentz factor~\cite{Chodorowski}. Therefore, for propagation
distances $\alt 100~{\rm Mpc}$, pair production from $^4$He can be
safely neglected. For $E \alt 10^{11}~{\rm GeV}$, photo-pion
production is also negligible because it has a threshold energy $\sim
145~{\rm MeV}$ in the nucleus rest frame. In this decade of energy
photo-disintegration is the dominant process for energy loss of $^4$He
nuclei: the peak of the GDR corresponds to photon energies of $27~{\rm
  MeV}$. With this dominance, we now exploit a complete analytic
treatment of the GZK energy loss using the simple form of our parametrization.

The interaction time $\tau_{\rm int}$ for a highly relativistic
nucleus propagating through an isotropic photon
background with energy $\varepsilon$ and spectrum
$dn(\varepsilon)/d\varepsilon$, is~\cite{Stecker:1969fw}
\begin{equation}
\frac{1}{\tau_{\rm int}} =
\frac{c}{2} \, \int_{\varepsilon_{\rm th}/2\gamma}^{\infty} \frac{1}{\gamma^2 \varepsilon^2} \
\frac{d n(\varepsilon)}{d \varepsilon}
\, d\varepsilon \, \int_{\e_{\rm th}}^{2\gamma \varepsilon} \varepsilon' \,
\sigma_A(\varepsilon') \, d\varepsilon' \, ,
\label{rate}
\end{equation}
where $\gamma \sim E/(A m_p)$ is
the Lorentz factor and $\sigma_A (\varepsilon')$ is the cross-section for
photo-disintegration by a photon of energy
$\varepsilon'$ in the rest frame of the nucleus. 
Inserting  (\ref{sigma}) into (\ref{rate}) we obtain
\begin{eqnarray}
\!\!\!\!\!\frac{1}{\tau_{\rm int}}  & \approx &  
\frac{ c\,  {\cal A} \, \varepsilon_0 }{2 \gamma^2}
\int_{\e_{\rm th}/2\gamma}^\infty \frac{d\varepsilon}{\varepsilon^2}\,\,
  \frac{d n(\varepsilon)}{d\varepsilon} \ \Theta (2 \gamma \varepsilon
  - \varepsilon_0 ) \nonumber \\
&  \approx & \frac{\sqrt{\pi} \, c\, \sigma_0 \, \varepsilon_0 \,
  (\varepsilon_0 -\varepsilon_{\rm th})\,
  \Gamma e^{\Gamma^2/2}}{\sqrt{2} \ \gamma^2}
\int_{\varepsilon_0/2 \gamma}^\infty \frac{d\varepsilon}{\varepsilon^2}\,\,
  \frac{d n(\varepsilon)}{d\varepsilon} .
\label{kk}
\end{eqnarray}
For the CMB,
\begin{equation}
\frac{d n(\varepsilon)}{d\varepsilon} =   \frac{1}{(\hslash
c)^3} \ \left(\frac{\varepsilon}{\pi} \right)^2 \
\left[e^{\varepsilon/T}-1 \right]^{-1} \,\,,
\label{nBE}
\end{equation}
and so (\ref{kk}) becomes
\begin{equation}
\frac{1}{\tau_{\rm int}}  \approx   \frac{\sigma_0 \, \varepsilon_0 \,
  (\varepsilon_0 - \varepsilon_{\rm th})\,
\Gamma\, e^{\Gamma^2/2} \, T}{\sqrt{2 \pi}\,\pi \ \hslash^3 c^2 \, \gamma^2} \,\,
\left| \ln \left(1 - e^{-\varepsilon_0/2 \gamma T}\right) \right| \,, 
\label{RBE}
\end{equation}
with  $T = 2.7255(6)~{\rm K}$~\cite{Fixsen:2009ug}.

Despite the computational convenience of the narrow width
approximation, a full calculation of the interaction time can be
achieved. The second integral in (\ref{rate}) can be calculated
exactly for the cross section (\ref{eq:shiftlognorm}) to give
\begin{eqnarray}
J(\varepsilon)  & = & \int_{\varepsilon_{\mathrm th}}^\varepsilon \varepsilon' \,
\sigma_A(\varepsilon') \, d\varepsilon' \nonumber \\ 
& = & \frac{{\cal A} }{2}\left[\varepsilon_{\rm th}\ {\rm
    erfc}\left(\frac{\Gamma^2-z(\varepsilon)}{\sqrt 2
      \Gamma}\right)+e^{3\Gamma^2/2}(\varepsilon_0-\varepsilon_{\rm
    th}) \right. \nonumber \\
& \times & \left.  {\rm erfc}\left(\frac{2\Gamma^2-z(\varepsilon)}{\sqrt 2
      \Gamma}\right)\right] \, .
\end{eqnarray}
For the CMB spectrum, (\ref{rate}) can be rewritten as 
\begin{equation}
\frac{1}{\tau_{\rm int}}=
\frac{c}{4\pi^2(\hbar c\gamma)^3}\int_{\varepsilon_{\rm th}}^\infty
\frac{J(\varepsilon)}{e^{\varepsilon/2\gamma T}-1}d\varepsilon.
\label{int-ultima}
\end{equation}
The integral in (\ref{int-ultima}) is solved numerically, allowing us
to obtain the present-day (redshift $z=0$) mean free path for $A=4$
nuclei travelling through the CMB with a Lorentz factor $\gamma$ as
\begin{equation}
  \lambda(\gamma)=4\pi^2 (\hbar c \gamma)^3\left(\int_{\varepsilon_{\rm th}}^\infty
    \frac{J(\varepsilon)}{e^{\varepsilon/2\gamma T}-1}d\varepsilon\right)^{-1}.\label{eq:mfp}\end{equation}
The mean free path is analogously calculated for the three other
models obtaining three functions $\lambda_k(\gamma)$, for $k\in\{{\rm
  KT, PSB, R}\}$, where KT stands for Karakula-Tkaczyk~\cite{Karakula:1993he}, PSB for
Puget-Stecker-Bredekamp~\cite{Puget:1976nz}, and R for
Rachen~\cite{Rachen}; see Appendix~\ref{app:1} 
for details. The PSB-model
has been the benchmark for the SimProp Monte Carlo code~\cite{Aloisio:2017iyh} whereas the
R-model is used by the CRPropa program~\cite{Batista:2016yrx}. In Fig.~\ref{fig:3} we show the mean free path for
$^4$He photo-disintegration on the CMB for the four considered models,
and the ratios $\lambda_k(\gamma)/\lambda(\gamma)$ for the three
models. In Fig.~\ref{fig:4}  we display  the relative error
$(\lambda-\lambda_{\rm NWA})/\lambda$ of the NWA as a function of energy.

In order to study the consequences that the different cross sections
have on particle propagation through the CMB, we study its
transmittance to $^4$He nuclei going through a given distance at a
given energy. We define $\mathcal T(\gamma,L)\equiv
e^{-L/\lambda(\gamma)}$ for the mean free path (\ref{eq:mfp}), and
$\mathcal T_k(\gamma,L)\equiv e^{-L/\lambda_k(\gamma)}$ for the other
three models. Since the model introduced in this paper provides the
smallest cross section, it will give the largest transmittance. To
study this, we define the relative transmittances $R_k(\gamma,L)\equiv
\mathcal T_k(\gamma,L)/\mathcal T(\gamma,L)$. The three ratios are
shown in Fig.~\ref{fig:5}. For a propagation distance of $3.5~{\rm
  Mpc}$, the transmission of the CMB for our cross-section model at
$10^{10.8}~{\rm GeV}$ is $\mathcal T\approx 0.11$.  Our calculations
also demonstrate that if e.g., there was a source a 3.5~Mpc and
deflections on the extragalactic magnetic field are small, the Earthly
$^4$He flux would be $35\%$ larger than the output of CRPropa
3~\cite{Batista:2016yrx} and $42\%$ larger than the output of SimProp
v2r4~\cite{Aloisio:2017iyh}. For a propagation distance of 4~Mpc, the
discrepancy increases as the Earthly fluxes would be $41\%$ and $49\%$
larger than those predicted by CRPropa 3 and SimProp v2r4,
respectively. Thus, even for CRPropa 3, which uses the best
among the older parameterizations, the differences introduced by a more careful accounting
of the $^4$He photo-disintegration cross section are significant. 

\begin{figure}
  \postscript{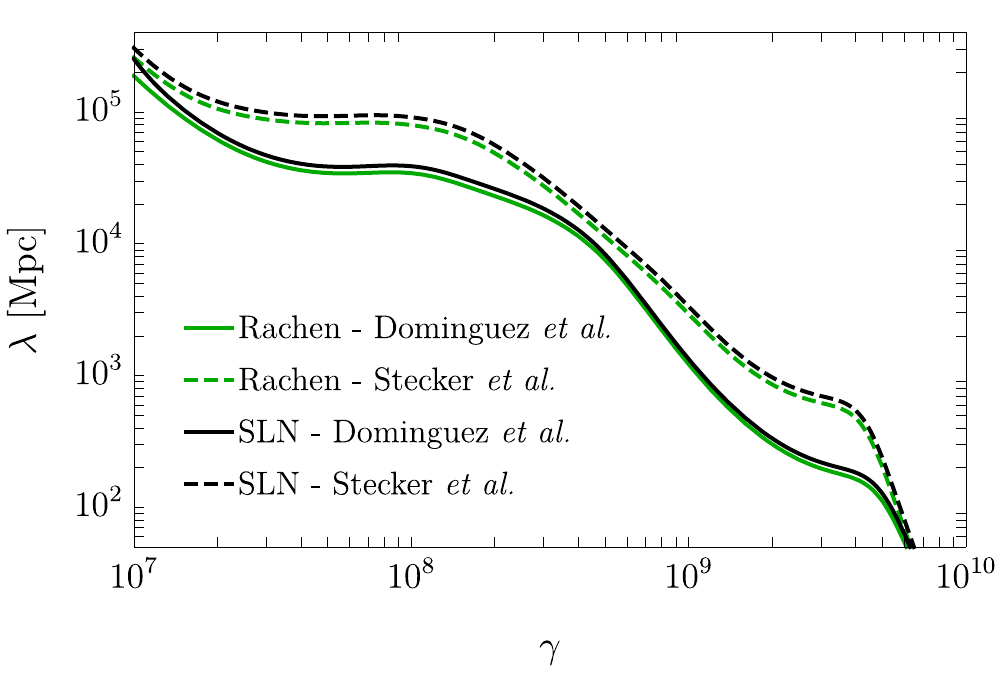}{0.9}  
  \caption{Photodisintegration mean-free-path of $^4$He on the
    IR photon background as estimated
    in~\cite{Dominguez:2010bv} and the lower limit derived in
    \cite{Stecker:2016fsg}. In the comparison we have used the
    photo-disintegration derived in this work and those obtained
    earlier by Rachen~\cite{Rachen}.}
\label{fig:6}
\end{figure}

For $\gamma \alt 10^{9.7}$ the dominant target photons are those of
the extragalactic background light. At present, the ambiguity in the
determination of infrared (IR) photon
background~\cite{Dominguez:2010bv,Gilmore:2011ks,Stecker:2016fsg} largely dominates
the uncertainties in the $^{4}$He mean-free-path.  This is illustrated
in Fig.~\ref{fig:6} where we show a comparison using the IR
estimates from~\cite{Dominguez:2010bv} and~\cite{Stecker:2016fsg}.

\section{Conclusion}

We have provided new parametrizations for the photo-disintegration
cross-section of nuclei with baryon number $A \leq 4$. In our fits we
included the most recent cross-section data both from the inclusive
measurement observing the analog of the giant dipole resonance in
$^4$He through the charge-exchange spin-flip {$^4$He} ($^7$Li,$^7$Be)
reaction and from measurements of exclusive two-body and three-body
processes: ${^4{\rm He}} \, (\gamma,p)\, {^3{\rm H}}$, ${^4{\rm He}}
\, (\gamma,n) \, {^3{\rm He}}$, and {$^4$He} $(\gamma, pn)$ {$^2$H}.
A comparison with previous estimates is displayed in Figs.~\ref{fig:1}
and \ref{fig:2}.

We have shown that existing Monte Carlo simulation codes for UHECR
propagation underestimate the predicted flux of $^4$He nuclei emitted
by sources in our cosmic backyard. For example, we demonstrated that
the mean free path  of $^4$He with $\gamma \sim 10^{10}$ increases
by more that 15\% with respect to previous estimates adopted as
benchmarks for Monte Carlo simulation codes of UHECR propagation. A
comparison of the different mean-free paths of $^4$He on the CMB for
relevant Lorentz factors is provided in
Fig.~\ref{fig:3}. Interestingly, the larger mean free path obtained in
our study implies that the {\it physical} survival probability of
$^4$He nuclei would be larger than predicted by existing event
generators.  For example, for $E \sim 10^{10.8}~{\rm GeV}$ and a
propagation distance of 3.5~Mpc, the $^4$He intensity would be $35\%$
larger than the output of CRPropa 3 program and $42\%$ larger than the
output of SimProp v2r4 program. A comparison of the increment in the
survival probability of $^4$He as a function of energy is exhibited in
Fig.~\ref{fig:5}.

As it is obvious, our finding have a direct impact on the possibility
that nearby starbursts could relate to the origin of cosmic-rays, what
we shall explore elsewhere.  It also provides a refreshing humble
perspective: basic nuclear physics can still significantly affect our
most common assumptions when imagining cosmic ray production sources.

\section*{Acknowledgments}

This work has been supported by the U.S. National Science Foundation
(NSF Grant PHY-1620661), the National Aeronautics and Space
Administration (NASA Grant 80NSSC18K0464), as well as by grants
AYA2015-71042-P, iLink 2017-1238, and SGR 2017-1383.

\appendix

\onecolumngrid

\section{Previous parametrizations of the giant dipole resonance}
\label{app:1}

In this Appendix we provide a brief description of the various cross-section models.

Karakula and Tkaczyk (KT) use a Breit-Wigner form  to model the peak
of the GDR  and fit the cross-section to a constant above 30~MeV,
\begin{equation}
\sigma^{\rm KT} _A(\varepsilon)=\left\{\begin{array}{l r} \sigma^{\rm
      KT}_0\,A \
   \frac{
     (\varepsilon\,\Gamma)^2}{(\varepsilon^2-\varepsilon_0^2)^2+(\varepsilon\Gamma)^2},&\varepsilon\leq\varepsilon^*
   \\
   A/8~{\rm
      mb},&\varepsilon>\varepsilon^* \end{array}\right. \,,
\end{equation}
with $\Gamma=8\,{\rm MeV}$, $\varepsilon_0=0.925\,A^{2.433}\,{\rm MeV}$, $\varepsilon^*=30\,{\rm MeV}$,  and $\sigma_0^{\rm KT}=1.45\,{\rm mb}$~\cite{Karakula:1993he} .

\begingroup
\renewcommand*{\thefootnote}{\alph{footnote}}
\begin{table}
\begin{center}
\caption{Parameters for the PSB cross sections. \label{tablaPSB}}
\begin{tabular}{ccccccc}
\hline
\hline
~~~~~~$A$~~~~~~&~~~~~~$i$~~~~~~&~~~~~~$\varepsilon_{{\rm th},i}$ (MeV)\footnotemark[1]
~~~~~~&~~~~~~$\varepsilon_{0,i}$
(MeV)~~~~~~&~~~~~~$\Delta_i$ (MeV)~~~~~~&~~~~~~$\xi_i$~~~~~~&~~~~~~$\zeta$~~~~~~ \\
\hline
\multirow{2}{*}{4}	&$1$&$20.2$&$27$&$12$&$0.47$&\multirow{2}{*}{$1.11$}\\
							&$2$&$27.2$&$45$&$40$&$0.11$&\\
\multirow{2}{*}{3}	&$1$&$6.6$&$13$&$18$&$0.33$&\multirow{2}{*}{$1.11$}\\
							&$2$&$6.6$\footnotemark[2]&$15$&$13$&$0.33$&\\
2&$1\footnotemark[3]$&$2.2$&$5$&$15$&$0.97$&---\\				
\hline
\hline
\end{tabular}
\end{center}
\end{table}
\footnotetext[1]{The source  often gives two energy thresholds
  corresponding to proton and neutron emission~\cite{Stecker:1998ib}. In our calculations we have taken the average value.}
\footnotetext[2]{The source does not provide this energy threshold~\cite{Stecker:1998ib}. Following~\cite{Skibinski:2003de}, we assume the energy threshold is similar to
  that of single nucleon emission.}
\endgroup

Puget, Stecker, and Bredekamp (PSB) also use a piecewise function
containing a Gaussian form (\ref{eq:norm}) around the peak of the GDR
and a constant above 30~MeV, with normalization given by the TRK
dipole sum rule~\cite{Puget:1976nz,Stecker:1998ib}.  PSB model the
total cross section as the sum of (at most) two contributions: single and multiple nucleon emission ($i=1,2$ respectively),
where
\begin{equation}
\sigma^{\rm PSB}_{A,i}(\varepsilon)=\left\{\begin{array}{l l}
    \xi_{i}\Sigma
    W_{i}^{-1}\exp\left(-\frac{2(\varepsilon-\varepsilon_{0,i})^2}{\Delta_i^2}\right),&\varepsilon_{{\rm
        th},i}\leq\varepsilon \leq\varepsilon^*\\
    \zeta\Sigma/(\varepsilon_{\rm
      max}-\varepsilon^*),&\varepsilon^*<\varepsilon\leq\varepsilon_{\rm
      max}\\ 0,&\varepsilon>\varepsilon_{\rm max}\end{array}\right. \,,
\end{equation}
and where
$\varepsilon^*=30\,{\rm MeV}$, $\varepsilon_{\rm max}=150\,{\rm
  MeV}$, with \begin{equation}
W_i=\Delta_i\sqrt{\frac{\pi}{8}}\left[{\rm
    erf}\left(\frac{\varepsilon^*-\varepsilon_{0,i}}{\Delta_i/\sqrt2}\right)+{\rm
    erf}\left(\frac{\varepsilon_{0,i}-\varepsilon_{{\rm
          th},i}}{\Delta_i/\sqrt2}\right)\right].\end{equation} 
The values of $\zeta$, $\xi_i$, $\varepsilon_{0,i}$ and $\Delta_i$ are taken from Table 1 of~\cite{Puget:1976nz}, and the threshold energies $\varepsilon_{{\rm th},i}$ are taken from Table 1 of~\cite{Stecker:1998ib}. These values are gathered here and shown in Table~\ref{tablaPSB} for each $A$ and $i$.

Rachen (R) uses two functional forms to parametrize the GDR
cross-sections of the different nuclei and processes including single ($i=1$)
or multiple ($i=2$) nucleon emission~\cite{Rachen}: {\it (i)}~the BP
form, and {\it (ii)} the function 
\begin{equation}{\rm Pl}(\varepsilon,\varepsilon_{\rm
    th},\varepsilon_{\rm
    max},\alpha)=\left(\frac{\varepsilon-\varepsilon_{\rm
        th}}{\varepsilon_{\rm max}-\varepsilon_{\rm
        th}}\right)^{\alpha(\varepsilon_{\rm max}/\varepsilon_{\rm
      th}-1)}\left(\frac{\varepsilon_{\rm
        max}}{\varepsilon}\right)^{\alpha\varepsilon_{\rm
      max}/\varepsilon_{\rm th}},\quad \varepsilon>\varepsilon_{\rm
    th} \,,\end{equation}
which has a maximum
at $\varepsilon_{\rm max}$ and a power law behavior both near threshold
and in the asymptotic limit; note that ${\rm Pl}(\e_{\rm th},\e_{\rm th}, \e_{\rm
max}, \alpha) = 0$  and $\lim_{\e \to \infty} {\rm Pl} (\e,\e_{\rm th}, \e_{\rm
max}, \alpha) \propto \e^{-\alpha}$. 

The  cross section for the two-body ${^4{\rm He}} \, (\gamma,p)\,
{^3{\rm H}}$ and ${^4{\rm He}} \,
  (\gamma,n) \, {^3{\rm He}}$ reactions is described by 
\begin{equation}
\sigma_{4,1}^{\rm R}=3.8\,{\rm mb}\, {\rm
  Pl}(\varepsilon,\varepsilon_{\rm th},\varepsilon_{\rm max},\alpha)
\,,\end{equation} with $\varepsilon_{\rm th}=19.8\,{\rm MeV}$,
$\varepsilon_{\rm max}=27\,{\rm MeV}$ and $\alpha=5$. The rest of the
processes are modelled using the BP form \begin{equation}
  \sigma_{A,i}^{\rm R}=\beta_{A,i}\,\sigma_{\rm
    BP}(\varepsilon,B_{A,i}) \,,\end{equation} 
where the nonzero coefficients are given in Table~\ref{tablaRachen}.

\begin{table}
\caption{Parameters for the Rachen cross sections. \label{tablaRachen}}
\begin{tabular}{cccc}
\hline
\hline
~~~~~~~~~~~~~$A$~~~~~~~~~~~~~&~~~~~~~~~~~~~$i$~~~~~~~~~~~~~&~~~~~~~~~~~~~$\beta_{A,i}$~~~~~~~~~~~~~&~~~~~~~~~~~~~$B_{A,i}$
(MeV)~~~~~~~~~~~~~\\
\hline
						$4$	&$2$&$1.4$&$26.1$\\
\multirow{2}{*}{$3$}	&$1$&$1.4$&$5.8$\\
								&$2$&$1.7$&$7.3$\\
						$2$	&$1$&$2.2$&$1.71$\\				
\hline
\hline
\end{tabular}
\end{table}

\end{document}